\documentclass[letterpaper,english,reprint, aps]{revtex4-1}
\usepackage[T1]{fontenc}
\usepackage[latin9]{inputenc}
\usepackage{babel}
\usepackage{amsmath}
\usepackage{amssymb}
\usepackage[unicode=true]
 {hyperref}

\makeatletter

\providecommand{\tabularnewline}{\\}

\makeatother

\begin{document}
\title{Effective conductivities of some multi-color, isotropic, regular tessellations}
\author{Clinton DeW.\,Van Siclen}
\email{cvansiclen@gmail.com}

\address{1435 W 8750 N, Tetonia, Idaho 83452, USA}
\date{15 January 2024}
\begin{abstract}
Algebraic expressions are found for the effective conductivities of
some infinite tessellations composed of conducting square, triangular,
or hexagonal tiles. A tessellation is further characterized by the
number $N$ of different colors (different tile conductivities) represented.
Tiles of a color are distributed randomly, and constitute an areal
fraction $1/N$ of the tessellation. The expressions take account
of the percolation threshold associated with the tile type. This generates
a series of expressions for the three-color case, that suggests this
approach gives lower bounds for the true effective conductivities.
\end{abstract}
\maketitle

\section{Introduction}

The effective properties of multi-component materials are of fundamental
and technological interest. Unfortunately these complex systems may
not always permit accurate calculation of property values. An illustrative
example is the effective conductivity $\sigma_{e}$ of a two-component,
two-dimensional (2D) system having a checkerboard morphology. As the
ratio $a/b$ of the two component conductivities increases, the current
flux due to a potential drop imposed across the checkerboard increasingly
concentrates at the corners of the component domains {[}\citealp{Soder},\citealp{Keller}{]}.
Those corners are infinitely sharp, thus requiring very high spatial
resolution to calculate a useful value for the effective conductivity.
However an algebraic approach gives the exact result, $\sigma_{e}=(ab)^{1/2}$.

This motivates a practical interest in finding algebraic expressions
for exact property values, or at least upper or lower bounds. Of course
the expressions must account for the particular symmetries and conditions
that characterize the system.

This paper considers the effective conductivity of regular tessellations:
those composed of square, triangular, or hexagonal tiles. The tessellations
have tiles of $N$ different ``colors'' (scalar conductivities).
Tiles of a color are distributed randomly and comprise areal fraction
$1/N$ of the statistically isotropic, infinite tessellation.

Thus the effective conductivity of a tessellation is a function $\sigma_{e}(a,b,c,d,..)$,
where the $N$ variables are conductivity values. Tiles of each conductivity
value make up an areal fraction $1/N$ of the tessellation.

Of particular interest are the $N=3$ tessellations, as those may
be considered by other methods {[}\citealp{Fel},\citealp{Bulgadaev03},\citealp{Bulgadaev04},\citealp{Barash}{]}.
For each tile type a series of algebraic expressions for the effective
conductivity is obtained.

The generic properties that the algebraic expressions must have are
\[
\sigma_{e}(k\:a,k\:b,k\:c,k\:d,...)=k\:\sigma_{e}(a,b,c,d,...)
\]
where the multiplier $k$ is any non-negative real number, and
\[
\sigma_{e}(a,a,a,a,...)=a.
\]

Particular conditions on the form of the expressions are that they
must be symmetric with respect to the $N$ colors; and that $\sigma_{e}=0$
when fewer than $X$ of the $N$ colors are conductive, where $X$
is the smallest integer that satisfies the relation $X/N>p_{c}$,
the value $p_{c}$ being the percolation threshold for the tile type;
and that the duality relation must be satisfied. That relation is
derived in the following section.

\section{Derivation of the duality condition}

Consider the two-component, 2D system $(\phi,a;1-\phi,b)$ that is
a square bond lattice in which a fraction $\phi$ of the bonds have
conductivity $a$ and the fraction $(1-\phi)$ of bonds have conductivity
$b$. The dual of this system is constructed by crossing each $a$
bond with an $a^{-1}$ bond and crossing each $b$ bond with a $b^{-1}$
bond. The dual system $(\phi,a^{-1};1-\phi,b^{-1})$ is again a 2D
square bond lattice. A straightforward calculation {[}\citealp{Straley}{]}
shows that the conductivity of the original system equals the resistivity
of its dual. Thus
\begin{equation}
\sigma_{e}(\phi,a;1-\phi,b)\times\sigma_{e}(\phi,a^{-1};1-\phi,b^{-1})=1.\label{eq:1}
\end{equation}
This derivation of the duality relation can be repeated for systems
with any number and distribution of components.

According to the principle of universality, any physical description
and properties of a structure in space are unaffected by how that
space is discretized. Thus the fraction $\phi$ in Eq.\,(\ref{eq:1})
may be replaced by the areal fraction $\phi$.

\section{Preliminary comments}

Note that a duality relation imposes a condition on the effective
conductivity of a multi-component system, but does not account for
the physical distribution and shapes of the component domains. As
a consequence, many algebraic expressions may satisfy that condition.

The simplest expression, that satisfies all conditions given in Sec.\,I,
has the form of a fraction, taken to a power chosen to ensure that
$\sigma_{e}$ has the unit of conductivity. The numerator is the sum
of the $N$ possible multiplications of $X$ colors (conductivities),
and the denominator is the sum of the $N$ possible multiplications
of $N-X$ colors (conductivities). The value $X$ is the smallest
integer that satisfies the relation $X/N>p_{c}$, and accounts for
the fact that $\sigma_{e}=0$ when there is no conducting path across
the infinite tessellation. This occurs when fewer than $X$ colors
are conductive.

An example is the expression for the effective conductivity of a three-color
tessellation composed of square tiles,
\begin{equation}
\sigma_{3}\left(a,b,c\right)=\frac{ab+bc+ca}{a+b+c}.\label{eq:2}
\end{equation}
Here the percolation threshold $p_{c}=0.592746$ so $X=2$. Note that
the subscript $3$ above indicates a three-color tessellation. This
notation ($\sigma_{N}$) will be used for the expressions that follow.

This expression for $\sigma_{3}$ satisfies the duality condition
\begin{equation}
\sigma_{3}\left(a,b,c\right)\times\sigma_{3}\left(a^{-1},b^{-1},c^{-1}\right)=1\label{eq:3}
\end{equation}
 since
\begin{equation}
\sigma_{3}\left(a^{-1},b^{-1},c^{-1}\right)=\frac{\left(a+b+c\right)/\left(abc\right)}{\left(ab+bc+ca\right)/\left(abc\right)}.\label{eq:4}
\end{equation}

As first noted by Dykhne {[}\citealp{Dykhn}{]},
\begin{equation}
\sigma_{3}\left(a,b,c\right)=c\label{eq:5}
\end{equation}
when $c=(ab)^{1/2}.$ This result is automatic for all expressions
$\sigma_{3}$ that satisfy the duality condition Eq.\,(\ref{eq:3}),
as it is obtained by simply multiplying both sides of Eq.\,(\ref{eq:3})
by $ab$.

Similarly, the relation
\begin{equation}
\sigma_{2}(a,b)=(ab)^{1/2}\label{eq:6}
\end{equation}
follows directly from the duality condition for two-component systems.

Another notational convention is useful below: $\sigma_{6}(a,b,c)$
means $\sigma_{6}(a,a,b,b,c,c)$, for example.

\smallskip{}

Applying this convention, the duality condition ensures that $\sigma_{N}(a,b)=(ab)^{1/2}$
for all even values of $N$; and that $\sigma_{N}(a,b,c)=c$ when
$c=(ab)^{1/2}$, for all values of $N$ that are multiples of $3$.

\section{Effective conductivities of tessellations with square tiles}
\begin{widetext}
Infinite tessellations composed of square tiles have the percolation
threshold $p_{c}=0.592746$. Thus the $N=6$ expression, having $X=4$,
is
\begin{equation}
\sigma_{6}(a,b,c,d,e,f)=\left[\frac{abcd+bcde+cdef+defa+efab+fabc}{ab+bc+cd+de+ef+fa}\right]^{1/2}\label{eq:7}
\end{equation}
which gives the three-component expression
\begin{equation}
\sigma_{6}(a,b,c)=\left[\frac{a^{2}b^{2}+ab^{2}c+b^{2}c^{2}+bc^{2}a+c^{2}a^{2}+ca^{2}b}{a^{2}+ab+b^{2}+bc+c^{2}+ca}\right]^{1/2}.\label{eq:8}
\end{equation}

Similarly the $N=9$ ($X=6$) and $N=12$ ($X=8$) expressions produce
the three-component expressions
\begin{equation}
\sigma_{9}(a,b,c)=\left[\frac{a^{3}b^{3}+a^{2}b^{3}c+ab^{3}c^{2}+b^{3}c^{3}+b^{2}c^{3}a+bc^{3}a^{2}+c^{3}a^{3}+c^{2}a^{3}b+ca^{3}b^{2}}{a^{3}+a^{2}b+ab^{2}+b^{3}+b^{2}c+bc^{2}+c^{3}+c^{2}a+ca^{2}}\right]^{1/3}\label{eq:9}
\end{equation}
\begin{equation}
\sigma_{12}(a,b,c)=\left[\frac{a^{4}b^{4}+a^{3}b^{4}c+a^{2}b^{4}c^{2}+ab^{4}c^{3}+b^{4}c^{4}+b^{3}c^{4}a+b^{2}c^{4}a^{2}+bc^{4}a^{3}+c^{4}a^{4}+c^{3}a^{4}b+c^{2}a^{4}b^{2}+ca^{4}b^{3}}{a^{4}+a^{3}b+a^{2}b^{2}+ab^{3}+b^{4}+b^{3}c+b^{2}c^{2}+bc^{3}+c^{4}+c^{3}a+c^{2}a^{2}+ca^{3}}\right]^{1/4}.\label{eq:10}
\end{equation}
\end{widetext}

\section{Effective conductivities of tessellations with triangular tiles}
\begin{widetext}
Infinite tessellations composed of triangular tiles have the percolation
threshold $p_{c}=0.5$. The $N=3$ ($X=2$) expression, and the $N=6$
($X=4$) expression, are identical to those for the tessellations
with square tiles. Thus the three-component expressions Eq.\,(\ref{eq:2})
and Eq.\,(\ref{eq:7}) apply to tessellations with triangular tiles
as well. However the $N=9$ ($X=5$) and $N=12$ ($X=7$) expressions
produce the three-component expressions
\begin{equation}
\sigma_{9}(a,b,c)=\left[\frac{a^{3}b^{2}+a^{2}b^{3}+ab^{3}c+b^{3}c^{2}+b^{2}c^{3}+bc^{3}a+c^{3}a^{2}+c^{2}a^{3}+ca^{3}b}{a^{3}b+a^{2}b^{2}+ab^{3}+b^{3}c+b^{2}c^{2}+bc^{3}+c^{3}a+c^{2}a^{2}+ca^{3}}\right]\label{eq:11}
\end{equation}
\begin{equation}
\sigma_{12}(a,b,c)=\left[\frac{a^{4}b^{3}+a^{3}b^{4}+a^{2}b^{4}c+ab^{4}c^{2}+b^{4}c^{3}+b^{3}c^{4}+b^{2}c^{4}a+bc^{4}a^{2}+c^{4}a^{3}+c^{3}a^{4}+c^{2}a^{4}b+ca^{4}b^{2}}{a^{4}b+a^{3}b^{2}+a^{2}b^{3}+ab^{4}+b^{4}c+b^{3}c^{2}+b^{2}c^{3}+bc^{4}+c^{4}a+c^{3}a^{2}+c^{2}a^{3}+ca^{4}}\right]^{1/2}.\label{eq:12}
\end{equation}
\end{widetext}

\section{Effective conductivities of tessellations with hexagonal tiles}
\begin{widetext}
Infinite tessellations composed of hexagonal tiles have the percolation
threshold $p_{c}=0.6962$. Thus a three-component tessellation will
not conduct unless all three components are conductive. Then the simplest
expression for the $N=3$ case is
\begin{equation}
\sigma_{3}(a,b,c)=\left[abc\right]^{1/3}.\label{eq:13}
\end{equation}
The $N=6$ ($X=5$) and $N=9$ ($X=7$) and $N=12$ ($X=9$) expressions
produce the three-component expressions
\begin{equation}
\sigma_{6}(a,b,c)=\left[\frac{a^{2}b^{2}c+ab^{2}c^{2}+b^{2}c^{2}a+bc^{2}a^{2}+c^{2}a^{2}b+ca^{2}b^{2}}{2(a+b+c)}\right]^{1/4}=\left[\frac{abc\left(bc+ac+ab\right)}{a+b+c}\right]^{1/4}\label{eq:14}
\end{equation}
\begin{equation}
\sigma_{9}(a,b,c)=\left[\frac{a^{3}b^{3}c+a^{2}b^{3}c^{2}+ab^{3}c^{3}+b^{3}c^{3}a+b^{2}c^{3}a^{2}+bc^{3}a^{3}+c^{3}a^{3}b+c^{2}a^{3}b^{2}+ca^{3}b^{3}}{a^{2}+a^{2}+ab+b^{2}+b^{2}+bc+c^{2}+c^{2}+ca}\right]^{1/5}\label{eq:15}
\end{equation}
\begin{equation}
\sigma_{12}(a,b,c)=\left[\frac{a^{4}b^{4}c+a^{3}b^{4}c^{2}+a^{2}b^{4}c^{3}+ab^{4}c^{4}+b^{4}c^{4}a+b^{3}c^{4}a^{2}+b^{2}c^{4}a^{3}+bc^{4}a^{4}+c^{4}a^{4}b+c^{3}a^{4}b^{2}+c^{2}a^{4}b^{3}+ca^{4}b^{4}}{a^{3}+a^{3}+a^{2}b+ab^{2}+b^{3}+b^{3}+b^{2}c+bc^{2}+c^{3}+c^{3}+c^{2}a+ca^{2}}\right]^{1/6}\label{eq:17}
\end{equation}
\end{widetext}

\section{Discussion and conclusions}

Contrary to the results in Secs.\,IV, V, and VI above, the expressions
$\sigma_{N}(a,b,c)$ for a tessellation type should be \textit{identical},
since they pertain to the same physical tessellation. Evidently the
conditions for their construction laid out in Sec.\,I are insufficient
to give a unique expression $\sigma_{N}(a,b,c)$. Besides $N$, the
2D systems considered here are characterized only by their color isotropy,
manifested as the color symmetry condition, and percolation threshold.

Regardless, it is interesting to note that the three-component expression
is not universal (identical for the three tessellation types). In
the case of the square and triangular tessellations, this is revealed
by considering higher-$N$ expressions for the conductivities, as
those more precisely locate the percolation threshold and thereby
differentiate those tessellations.

For each tessellation type, a \textit{series} of expressions for $\sigma_{N}(a,b,c)$,
all satisfying the duality condition Eq.\,(\ref{eq:3}), was obtained.
Arguably, more-accurate conductivity values are obtained with increasing
$N$ value, allowing the ratio $X/N\rightarrow p_{c}$. To characterize
an expression for the purpose of comparison, the volume integral
\begin{equation}
I_{N}=\int_{a=0}^{a=1}\int_{b=0}^{b=1}\int_{c=0}^{c=1}\sigma_{N}(a,b,c)\,da\,db\,dc\label{eq:17-1}
\end{equation}
is calculated {[}\citealp{Math}{]}. Note that expanding the upper
limits by a factor of $10$ increases the result by a factor of $10^{4}$.

In effect, this integral sums all possible values of the expression
$\sigma_{N}(a,b,c)$. Of those, values of $\sigma_{N}(a,b,\sqrt{ab})$
are certainly correct.

\begin{table}[h]
\caption{$I_{N}$ values for the three-component tessellations}

\begin{tabular}{|c|c|c|c|}
\hline 
$N$ & triangular & square & hexagonal\tabularnewline
\hline 
\hline 
3 & \ 0.439772\  & \ 0.439772\  & \ 0.421875\ \tabularnewline
\hline 
6 & 0.446654 & 0.446654 & 0.425658\tabularnewline
\hline 
9 & 0.457265 & 0.452915 & 0.432054\tabularnewline
\hline 
\ 12~ & 0.462834 & 0.458075 & 0.438157\tabularnewline
\hline 
\end{tabular}
\end{table}

Table I contains the $I_{N}$ values for the three-component expressions.
Those values are seen to increase with $N$, which suggests that the
$\sigma_{N}(a,b,c)$ expressions give lower bounds for the true conductivities
of the isotropic, infinite tessellations. In addition, a comparison
of the $I_{N}$ values for the three tessellation types suggests that
the hexagonal tessellation is less conductive, for a set $\{a,b,c\}$
of component conductivities, than the other two types.
\begin{acknowledgments}
I thank Professor Indrajit Charit (Department of Nuclear Engineering
\& Industrial Management) for arranging my access to the resources
of the University of Idaho Library (Moscow, Idaho).
\end{acknowledgments}

\end{document}